\documentclass[conference]{IEEEtran}
\IEEEoverridecommandlockouts

\usepackage{cite}
\usepackage{xspace}
\usepackage{amsmath,amssymb,amsfonts}
\usepackage{tabularx}
\usepackage{enumitem}
\usepackage{graphicx}

\newcommand{\mtcpa}{\operatorname{tCPA}}
\newcommand{\pcvr}{\operatorname{pCVR}}
\newcommand{\offset}{\textsc{Offset}\xspace}
\DeclareMathOperator{\recloss}{RecLoss}
\DeclareMathOperator{\code}{Code}
\DeclareMathOperator{\expect}{\mathbb{E}}
\newcommand{\bv}[1]{\mathbf{#1}}

\newcommand{\logloss}{\textit{logloss}\xspace}
\newcommand{\Comment}[1]{}

    \newcommand{\ariel}[1]{}
    \newcommand{\yohay}[1]{}
    \newcommand{\alex}[1]{}
\Comment{
    \newcommand{\ariel}[1]{\noindent{\textcolor{purple}{\{{\bf Ariel:} {\em #1}\}}}}
    \newcommand{\yohay}[1]{\noindent{\textcolor{blue}{\{{\bf Yohay:} {\em #1}\}}}}
    \newcommand{\alex}[1]{\noindent{\textcolor{magenta}{\{{\bf Alex:} {\em #1}\}}}}
}

\begin{document}

\title{Improving conversion rate prediction via self-supervised pre-training in online advertising}

\author{\IEEEauthorblockN{Alex Shtoff}
\IEEEauthorblockA{\textit{Yahoo Research}\\
Haifa, Israel \\
alex.shtoff@yahooinc.com}
\and
\IEEEauthorblockN{Yohay Kaplan}
\IEEEauthorblockA{\textit{Yahoo Research}\\
Haifa, Israel \\
yohay@yahooinc.com}
\and
\IEEEauthorblockN{Ariel Raviv}
\IEEEauthorblockA{\textit{Yahoo Research}\\
Haifa, Israel \\
arielr@yahooinc.com}
}

\IEEEoverridecommandlockouts
\IEEEpubid{\makebox[\columnwidth]{979-8-3503-2445-7/23/\$31.00 ©2023 IEEE\hfill}
\hspace{\columnsep}\makebox[\columnwidth]{ }}

\maketitle

\begin{abstract}
The task of predicting conversion rates (CVR) lies at the heart of online advertising systems aiming to optimize bids to meet advertiser performance requirements. Even with the recent rise of deep neural networks, these predictions are often made by factorization machines (FM), especially in commercial settings where inference latency is key. These models are trained using the logistic regression framework on labeled tabular data formed from past user activity that is relevant to the task at hand. 

Many advertisers only care about click-attributed conversions, which are conversions that occurred after a user has clicked on an ad. A major challenge in training models that predict conversions-given-clicks comes from data sparsity - clicks are rare, conversions attributed to clicks are even rarer. However, mitigating sparsity by adding conversions that are not click-attributed to the training set impairs model calibration, causing the mean prediction to no longer converge to the actual CVR. Since calibration is critical to achieving advertiser goals, this is infeasible.

In this work we use the well-known idea of self-supervised pre-training, and use an auxiliary auto-encoder model trained on all conversion events, both click-attributed and not, as a feature extractor to enrich the main CVR prediction model. Since the main model does not train on non click-attributed conversions, this does not impair calibration. We adapt the basic self-supervised pre-training idea to our online advertising setup by using a loss function designed
for tabular data, facilitating continual learning by ensuring auto-encoder stability, and incorporating a neural network into a large-scale real-time ad auction that ranks tens of thousands of ads, while conforming to the strict latency constraints, and without incurring a major engineering cost. We evaluate our approach and show improvements both offline, during training, and in an online A/B test. Following its success in A/B tests, our solution is now fully deployed to the Yahoo native advertising system, and its impact is measured in millions of dollars annually.
\end{abstract}


\section{Introduction}

\begin{figure}[h]
    \centering
    \includegraphics[width=.9\columnwidth]{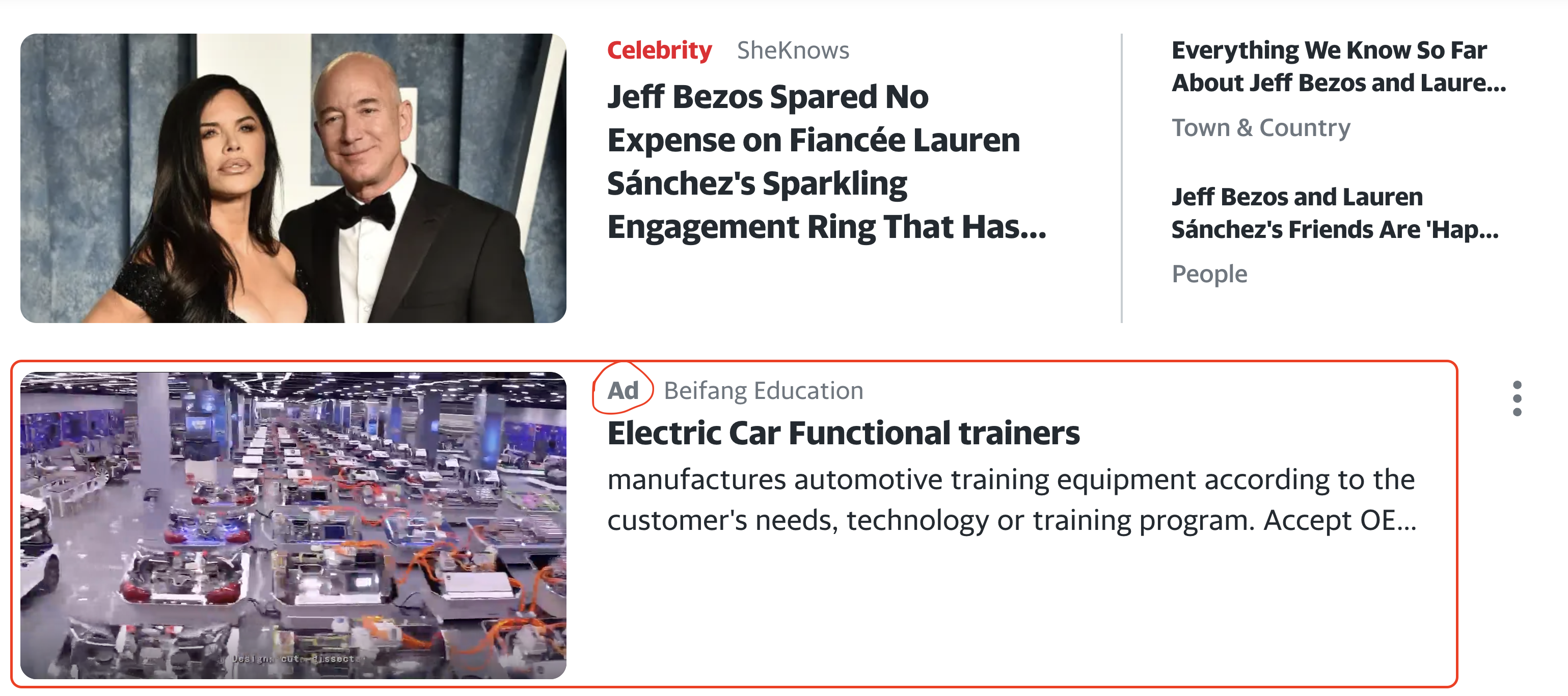}
    \caption{A native ad on Yahoo homepage that resembles the surrounding content.}
    \label{fig:native_ad}
\end{figure}

The Yahoo native ads marketplace serves users with native ads that resemble the surrounding content (see Figure \ref{fig:native_ad}). It shows billions of daily impressions with a run-rate of several hundreds of millions USD each year. In order to rank native ads for incoming users and their specific context according to the cost-per-click price type, the expected revenue of each ad is computed as a product of the advertiser's bid and the predicted click probability. The click probability is produced by a feature enhanced collaborative filtering algorithm called \offset \cite{aharon2013off}.

Advertiser bids, that should reflect the value of an ad click to the advertiser, are either specified manually, or chosen programmatically according to several strategies designed to meet various advertiser goals. At the heart of programmatic bid optimization lie \offset models that compute, for each ad, the predicted conversion rate (pCVR) given a click for a given user and a context. For example, one of the strategies allows advertisers to specify the sum they are willing to pay for a conversion, also known as the target cost per acquisition (tCPA). The bid, in that case, is 
\begin{equation*}
\operatorname{bid} = \pcvr \cdot \mtcpa.
\end{equation*}
Clearly, calibration of the $\pcvr$ model, in the sense that its average prediction approximates the true CVR on any traffic segment, is crucial. Consistent over-prediction causes over-bidding, which results in the advertisers paying more than they desire per conversion. Under-prediction causes under-bidding, which results in the advertisers winning less auctions and paying less for the ones they won, meaning the advertisers are losing exposure, and we are losing revenue.

\offset is a variant of a factorization machine \cite{rendle2010factorization} that is trained using the logistic regression framework involving positive and negative events. \offset trains incrementally on an infinite stream of data, with periodic checkpoints for hyper-parameter tuning \cite{aharon2017adaptive} and deployment to the production environment. The incremental training methodology was adopted to make sure that the delay between an event happening and a model trained on that event being deployed to production is as small as possible. 
In many systems \cite{debt, trends, zhang2020retrain}, including ours, model freshness is of paramount importance for the performance of content recommendation systems deployed in a changing environment.

Training CVR prediction model poses many key challenges, one of which is the data sparsity issue. Accurate models require a large amount of data, but the required amount of data when training conversion models is not always available. For example, typically a few percents of the impressions result in clicks, and a few percents of those result in a conversion. Thus, the number of clicks and conversions for training is quite small, and the coverage of various feature combinations, such as "25-30 years old people using a mobile device", is even smaller.

An initial idea that might come to mind is enriching the training set with additional data, such as conversions that are not attributed to clicks, or impression data in addition to clicks. However, since we require the model for the regression task of CVR prediction, rather than a classification task, it is unclear how the labels of such events should be assigned to produce a calibrated model.

To avoid feeding the main CVR prediction model with data that may impair calibration, we use the celebrated approach of self-supervised pre-training. We pre-train an auto-encoder model \cite{modular_learning_ballard,kramer1991nonlinear} on \emph{all} conversions, both related and unrelated to the task at hand. Then, we train the main CVR prediction model \emph{only} on the relevant data while using the encoder's output, which we refer to as the \emph{code}, as an additional feature, meaning that the we attach task-specific prediction layers on top of the encoder.  Since these layers are trained only on the task-specific data, no bias is introduced. Moreover, since the code is low-dimensional, training a low dimensional model is potentially immune to data sparsity. 

The objective of our design is to provide the CVR prediction model with valuable information about user conversion patterns available in the additional data present in non click-attributed conversions, without accessing it directly. The results of the online and offline experiments in this paper support our thesis by demonstrating improved performance of our CVR prediction model, both offline and in an online A/B test.

The setting of online advertising and incremental training pose unique challenges that require adapting the standard self-supervised pre-training methodology. Hence, we present several design decisions for tailoring the methodology towards a system performing auctions under strict latency constraints, driven by a factorization machine variant that is incrementally trained on tabular data. Concretely, the decisions include the choice of a loss function, careful feature selection, and a model architecture that is driven by a set of criteria and metrics designed to evaluate an incrementally training auto-encoder.

To summarize, the main contributions of our paper are:
\begin{itemize}
    \item An application of self-supervised pre-training to utilize additional data from conversion events that a CVR model cannot access directly during training.
    \item A set of techniques for training and using auto-encoders with incrementally trained factorization machine variants that are used in ranking under strict latency constraints.
\end{itemize}

\section{Related work}
Our work primarily addresses the task of conversion prediction\cite{zhang2014optimal}. This task is similar to click prediction, the 'core' task of online ad models, and often uses similar approaches. Its distinct challenges, such as the click-to-conversion delay, the difficulty in attribution to a specific event, and the attribution gap between different platforms and advertisers have been tackled in \cite{chapelle2014modeling,mahdian2007pay,shao2011data,lu2017practical} and many subsequent works.

In this work we adapt a self-supervised pre-training technique to the task of conversion rate prediction in a strict latency constraints settings. Self-supervised pre-training is the technique that leverages self-supervised learning \cite{mikolov2013distributed,devlin2018bert} to generate feature representations, typically using a deep neural network, that are (somewhat) agnostic of the overall prediction task and subsequently injecting them in some fashion to the dedicated predictive model. In previous works this main model has also been some type of deep-neural networks model \cite{sun2019bert4rec,yuan2020parameter, zhang2022keep}. In \cite{ni2018perceive} a universal user representation is learned from multiple forms of data and then applied to multiple predictive models. \cite{zhou2020s3} considers a sequential ordering of predictive tasks and injects representations learned from multiple tasks into various points of a self-attentive network. \cite{hu2018conet,yuan2019darec} suggest ways of taking the standard deep learning predictive flow (i.e., embedding layer into cross layers into predictive output) and injecting intermediate output of one task's model into the other (and vice versa) to boost the performance of both.

In contrast to recent works that tackle this task using deep neural networks, our underlying model is a variant of a factorization machine \cite{rendle2010factorization}, specifically \offset\cite{aharon2013off,aharon2017adaptive}. This is also the underlying model for most prediction tasks in Yahoo's native advertising network.

\section{Our approach}
In this section we describe the techniques we use to integrate an auto-encoder into the CVR prediction models in a way that facilitates incremental training, and enables real-time ad auctions under strict latency constraints.

\subsection{Training a CVR model with an autoencoder}
Like many probability-based regressions models, \offset models compute their output (in this case, a $\pcvr$) with the \textit{sigmoid} function:
\[
\pcvr(\Omega)=\frac{1}{1+e^{-f(\Omega)}}
\]
where $\Omega$ is the user,ad and contextual properties of the given event. The $f$ used by \offset can broken into:
\begin{equation}\label{eq:offset_score}
f(\Omega) = \langle \bv{U}, \bv{A} \rangle + \sum_{i\in S}w_i x_i + b 
\end{equation}
where $\bv{U}$ is a latent representation vector of the user and contextual features, $\bv{A}$ is a latent representation of the ad features, $S$ is a set of potential additional features, $x_i$ is an indicator the feature $i$, $w_i$ is the learned weight for feature $i$ and $b$ is a global bias. The continual model training process is responsible for generating the latent representation for all possible combinations of user,contextual and ad features as well as the various $w_i$'s and the global bias term. For more on how \offset constructs the various latent vectors from $\Omega$ and its parameters please see \cite{aharon2013off} and \cite{aharon2017adaptive}. $S$ is used for features that aren't separable between the user and ad, such as how often and how recently this user has seen this ad, see \cite{aharon_2019_sfc} for more. The training process uses the \logloss function as its error function:
\[
\mathcal{L}(\Omega,y,t)=-(1-y)\log\left(1-\pcvr(\Omega)\right) -y \log \pcvr(\Omega)
\]
where y is an indicator for a positive event (in our case, click attributed conversion)

We quickly note that this form of computation has benefits for the ability to handle real-time inference. Each ad has it pre-computed latent representation, while each user has a singular latent vector (calculated at inference time).

\begin{figure*}
  \centering
  \includegraphics[width=.9\textwidth]{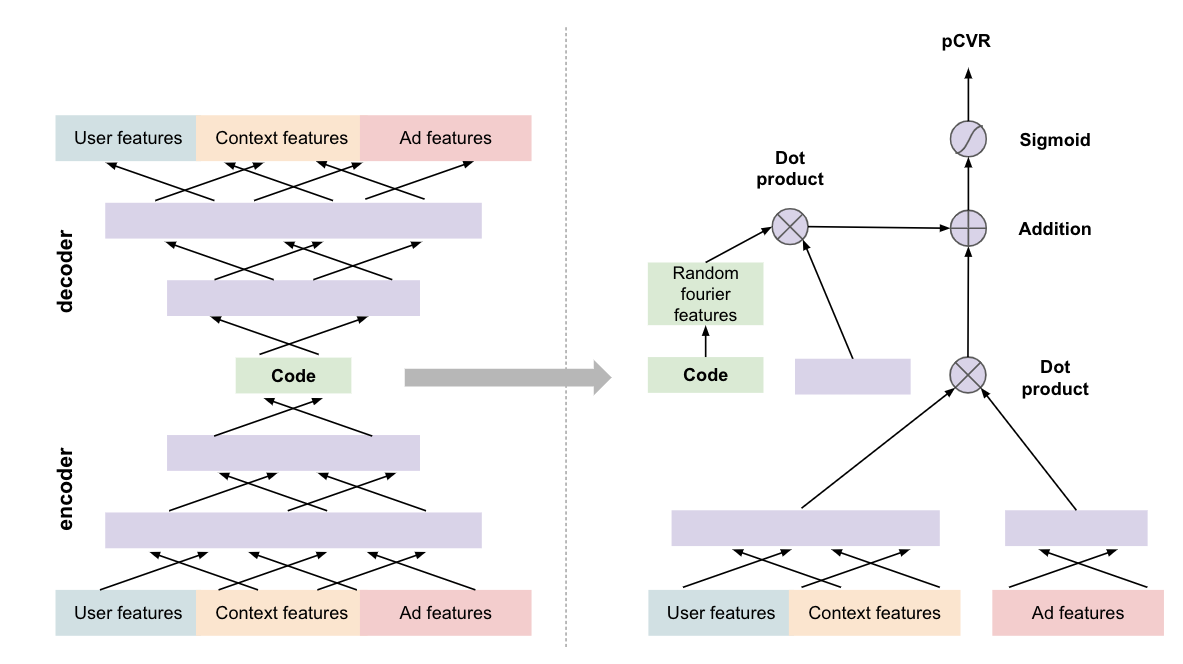}
  \caption{Schematic illustration of our self-supervised pre-training framework. On the left - the auto-encoder that is trained on all conversions, both click-attributed and not. On the right - the CVR prediction model that is trained on clicks and click-attributed conversions, and uses the encoder's code for its training data as an additional feature.}
  \label{fig:system_architecture}
\end{figure*}

We want to introduce the code generated by the auto-encoder into this computation. Since the code depends on both user and ad features, it makes sense to include it in the same way that $S$ is used, so as not to intrude on the portions of the computation that are user/ad isolated. So the resulting $f$ is 
\begin{equation}\label{eq:modified_offset_model}
f(\Omega) = \langle \bv{U},\bv{A} \rangle + \sum_{i\in S} w_i x_i + b + \langle \bv{W}, C(\Omega) \rangle
\end{equation}
where $C(\Omega)$ is the code associated with a specific event, and $\bv{W}$ is vector of weight learned by the model as part of the training process. The resulting system is illustrated in Figure \ref{fig:system_architecture}. Learning a linear function of the code is our way to re-use the existing training and serving infrastructure that is built for \offset models, that already include a linear term. Although at first glance it might look like having a limited representation power, we show in later sections how we significantly enhance it using a classical method in machine learning.

So a single training interval of the model does the following:
\begin{enumerate}
    \item in interval $t$ - load previous \offset model $M_{t-1}$ and previous encoder $Enc_{t-1}$
    \item For each event $\Omega$, calculate the code $C=Enc_{t-1}(\Omega)$ 
    \item \textit{(CVR training)} adapt model parameters (including $\bv{W}$) using SGD according to $f(\Omega,C)$ using \textit{logloss} error
    \item \textit{(Autoencoder trainig)} adapt parameters of the auto-encoder, whose encoder is $Enc_t$.
\end{enumerate}

\subsection{Architectural challenges}
Despite the architecture's apparent simplicity and ease of integration into an existing serving system, it poses several challenges. We describe them here, and present our solution in the following sub-sections. The first challenge stems from incremental training. The basic premise is that the model has long-term memory, and its parameters at interval $t-1$ are a good initialization point for training at interval $t$. Therefore, the codes $C(\Omega)$ need to be stable, in the sense that the expectation $\mathbb{E}_{\Omega} [Enc_{t-2}(\Omega) - Enc_{t-1}(\Omega)]$ is small. Otherwise, the premise is broken, since the vector $W$ that \offset learned in the previous interval encodes little information that is useful at interval $t$. 

Second, the expressive power of the linear function $\bv{W}^T C(\Omega)$ is weak. Indeed, typically the prediction layers on top of an auto-encoder is, by itself, a neural network of several layers. This is because the separation between positive and negative samples in the encoder's latent space is often non-linear.

The final challenge comes from our need to perform fast real-time auctions. In our system, the time it takes to use a neural network for every item in the auction introduces latency that is orders of magnitude beyond our latency constraints. 

\subsection{Auto-encoder architecture and design} \label{sec:autoencoder_design}
Suppose our data-set comprises of $C$ categorical columns where column $i$ has one of $n_i$ possible values. The encoder's first layer is an embedding layer - each column has an embedding table of dimension $n_i \times d$. Next, the embedding vectors are concatenated to form a vector of length $C \cdot d$, which is then passed to a multi-layer perceptron (MLP) network to produce the code. The hidden layers use ReLU activations, while the layer that produces the code uses a $\tanh$ activation to produce code that is bounded in the unit box.

\begin{figure}
    \centering
    \includegraphics[width=.9\columnwidth]{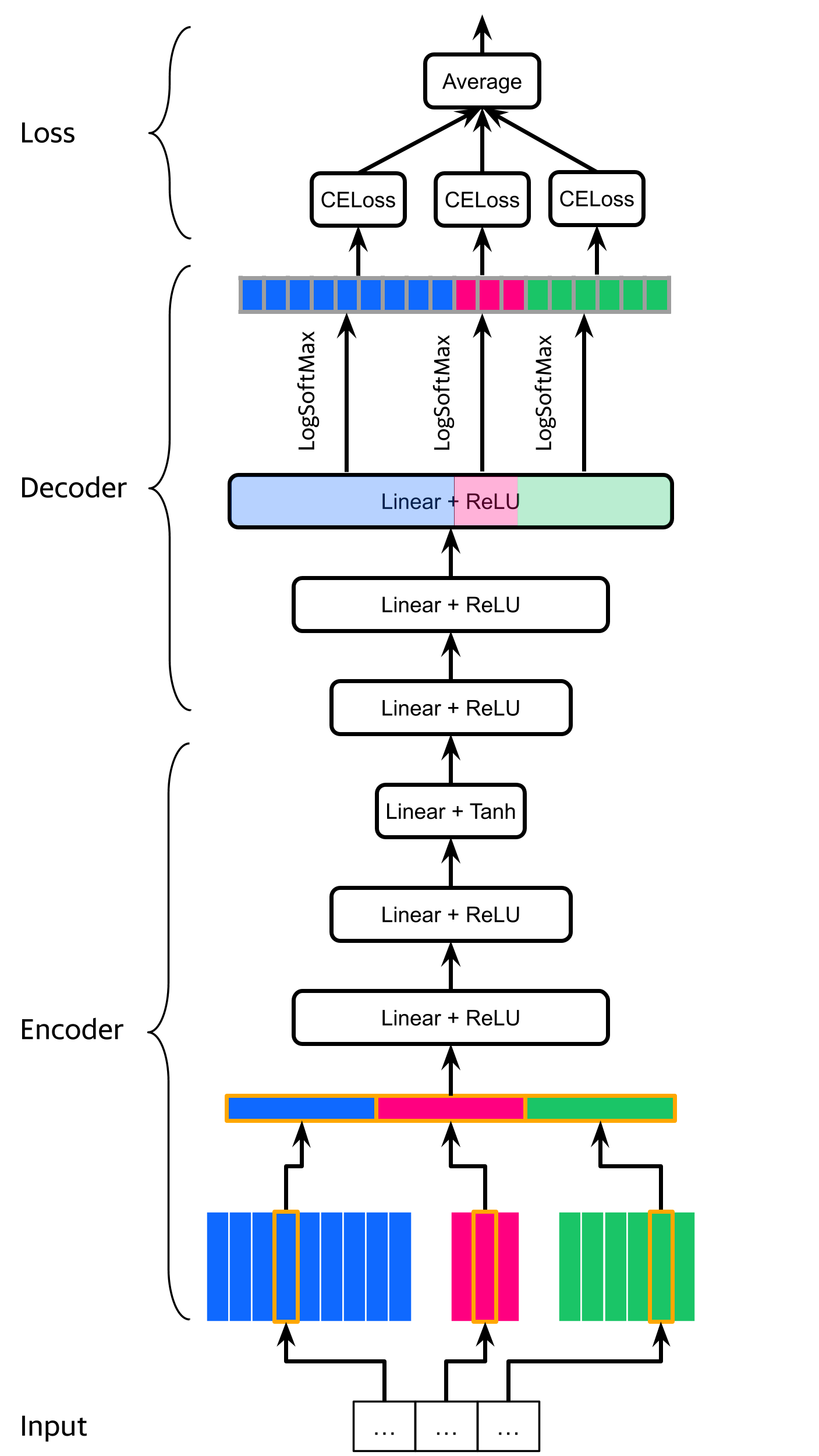}
    \caption{Auto-encoder model architecture. For clarity, illustrated with $C = 3$ tabular columns. The input is used to choose embeddings, that are concatenated, and fed to a multi-layer perceptron to produce the code. The decoder, in turn, is a multi-layer perceptron that acts as a set of classifiers with cross-entropy losses, that are averaged to obtain the output.}
    \label{fig:autoencoder_arch}
\end{figure}

Since the encoder's input is composed of categorical features, we treat the decoder as a set of multi-class classifiers with a classifier of $n_i$ classes for every column. To that end, it consists of an MLP that transforms the code to a vector of size $\sum_{i=1}^C n_i$, that we treat as $C$ blocks of size $n_i$ each. The last layer applies a log-softmax activation to each block to produce valid multi-class logits. The reconstruction loss is, naturally, the average of cross-entropy losses applied to each block. Our architecture is illustrated in Figure \ref{fig:autoencoder_arch}. We denote the loss of auto-encoder $M$ on sample $\Omega$ as $\recloss(M, \Omega)$, and the code it produces by $\code(M, \Omega)$. We note that the na\"{\i}ve method of reconstructing a concatenation of one-hot encoded feature indicators using the L2 loss did not result in any improvement of the downstream CVR model.

Just like \offset, our auto-encoder is trained in an online manner on data sub-divided into intervals $D_1, D_2, \dots$, training on each event only once. The process produces a sequence of auto-encoder models $M_1, M_2, \dots$. On our data, the optimizer that produced the best results was stochastic gradient-descent with momentum \cite{sgd, momentum}.  We found the online paradigm to produce satisfactory results, and our use of the paradigm is driven both by the practical success of \offset and theoretical results about generalization of online learning \cite{online_learning_generalization}. We note that in the online learning paradigm there is no test set separate from the training set, and instead each sample is first evaluated and then trained upon. 

The architecture we describe is the result of a process that was driven by empirical evaluation of several metrics that represent its quality in terms of learning data patterns, stability, and generalization. We gradually increased the complexity of the encoder until we found its performance to be satisfactory, while meeting the strict latency constraints. Beyond embedding, code, and hidden layer dimensions, we also tried different ways of combining the column embeddings: addition, product, sum of pairwise component-wise products (\`{a} la FM), sum of pairwise outer product matrices, and concatenation. The best results according to the metrics described below were obtained with concatenation.

We designed our model with several desired properties in mind. Below, we describe these properties and the metrics we used to measure the model's performance against them.
\paragraph{Small reconstruction loss} We aim at a reconstruction loss that is much smaller than that of uniformly distributed columns:
\[
\recloss_t \equiv \expect_{\Omega \sim D_t}[\recloss(M_t, \Omega)] \ll \frac{1}{C} \sum_{i=1}^C \ln(n_i).
\]
\paragraph{Stability} The code produced by the encoder of the current interval should be close to the one produced by the encoder of the previous interval. Therefore, the following quantity should be as small as possible:
\[
\mathrm{Diff}_t = \expect_{\Omega \sim D_t} [\| \code(M_t, \Omega) - \code(M_{t-1}, \Omega) \|].
\]

\paragraph{Interval generalization} Previous interval's auto-encoder should generalize well to the samples in the current interval, and the following quantity should be close to 1:
\[
\mathrm{Gen}_t = \frac{\expect_{\Omega \sim D_t}[\recloss(M_t, \Omega)]}{\expect_{\Omega \sim D_t}[\recloss(M_{t-1}, \Omega)]}.
\]

\paragraph{Learning meaningful patterns}  The encoder should reconstruct real data well, and random data badly. The random data-set $R_t$ is of the same length as $D_t$, but with column values chosen uniformly at random from the column's dictionary. The reconstruction loss on $R_t$ is denoted by $\recloss^R_t \equiv \expect_{\Omega \sim R_t}[\recloss(M_t, \Omega)]$. The random loss ratio, defined by
\begin{equation*}
\mathrm{RandRatio}_t = \frac{\recloss_t}{\recloss^R_t},
\end{equation*}
should be close to zero, preferably $<10^{-2}$ - the reconstruction of real data should be \emph{orders of magnitude} better than that of random data.

We would like to point out that interval generalization $\mathrm{Gen}_t$ is tightly related to stability $\mathrm{Diff}_t$. In fact, generalization implies stability. Intuitively, if the model from the last interval generalizes well to the current interval's data, then its loss gradients w.r.t the data of the current interval are small, and its parameters will not change significantly. In the evaluation section we show that this phenomenon happens in practice. 

We observed that the most profound positive effect on all of the above metrics was obtained from embedding concatenation, rather than other strategies of handling the embedding vectors, and from increasing the embedding dimension. Increasing the number of hidden layers beyond a certain amount had little effect, and so was increasing their dimension. At first the above seems to be in contrast to classical machine-learning, where increasing model complexity results in over-fitting and reduces generalization. However, it is known \cite{overparam1, overparam2, overparam3} that over-parameterized networks, which are networks that are complex enough to achieve near zero training loss, generalize well. As is apparent in our evaluation section, our auto-encoder's training loss is indeed close to zero. This is despite the fact that it is quite small - we have 7 categorical columns, the embedding dimension is 20, the code dimension is 12, and the total number of parameters of our encoder is roughly $10^5$.

\subsection{Nonlinear separability}
Having learned an encoder, there is no guarantee that a linear function can differentiate well between positive and negative samples in the encoder's embedding space. Classical machine learning methodology suggests using kernel methods \cite{kernel1, kernel2, kernel3} as the go-to technique for creating linear models that represent non-linear functions. However, kernel methods require having the entire training set in advance, and thus do not fit incremental training. As a remedy, we use random Fourier features (RFFs) \cite{rff} as an approximation of the Gaussian radial basis function kernel. This technique suits incremental training well, since it dictates a simple formula for transforming the raw $\code(\Omega)$ into the feature vector $C(\Omega)$ in Equation \eqref{eq:modified_offset_model} with randomly generated matrix $\bv{P}$ and vector $\bv{q}$:
\[
C(\Omega) = \cos(\bv{P} \code(\Omega) + \bv{q}).
\]

To verify that RFFs have significantly more representation power, we ran a simple experiment of training a linear CVR model based on the encoder's code, with and without RFFs. We observed that RFFs provide a substantial improvement of 15\% in \logloss and 3.6\% in AUC. The model improves as the number of rows of $\bv{P}$ increases, until it reaches a plateau at 200 rows. This means that computing $C(\Omega)$ requires an additional multiplication by by a $200 \times 12$ matrix, which does not incur any significant latency cost.

\subsection{Latency constraints}
During an ad auction we rank tens of thousands of ads in a matter of milliseconds. Neural network inference for every ad is prohibitive, because it introduced a significant latency cost far beyond the acceptable range in our system. The above is true even for small encoders of a few tens of thousands of parameters. In theory, if we had only user features in the auto-encoder, we could use the code only \emph{once} per auction. However, an auto-encoder that was trained without features of the ads that were responsible for conversions did not improve the downstream CVR model. 

To avoid inference for every ad, we use only high-level ad taxonomy features in our auto-encoder. In our system, we have a two-level taxonomy hierarchy - the first level breaks down ads by high-level categories, such as \textit{Finance} or \textit{Tourism}, and the second level by a few dozen categories. Hence, during ranking, we need to compute the product $W^T C(\Omega)$ from Equation \eqref{eq:modified_offset_model} only once for every low-level category, and add the result to the score $f(\Omega)$ we compute for every ad according to its taxonomy.

\section{Evaluation}
We performed several experiments on our CVR models that predict the probability of a conversion given a click. For training our CVR prediction model, we treat clicks as negative events, and conversions attributed to clicks on our properties as positive events. In addition, we also have conversions that are attributed to ad views. All models train on daily intervals.

In this section we first corroborate our thesis that directly adding view-attributed conversions to the training set produces a less accurate model. Then, we evaluate our auto-encoder according to the metrics in Section \ref{sec:autoencoder_design}. Next, we show that equipping a CVR model with an auto-encoder improves the model's accuracy during training. Then, we verify that the incurred computational cost has no visible effect on the auction latency, and demonstrate the improved performance of the model on real traffic via an A/B test against the production model.

\subsection{Training on view-attributed conversions}
The paper's premise is that training on conversions that are not attributed to clicks, in addition to those that are attributed to clicks, degrades model accuracy. We trained a model that includes view-attributed conversions as positive events, in addition to the relevant events. In both models, the \logloss was evaluated only for clicks and conversions attributed to clicks. The \logloss achieved by the model that includes view-attributed conversion is between 3\% and 7\% higher (depending on the day) than that of the model which does not include them, which makes such a model much \emph{worse}.

\Comment{
\begin{figure}
    \centering
    \includegraphics[width=.7\columnwidth]{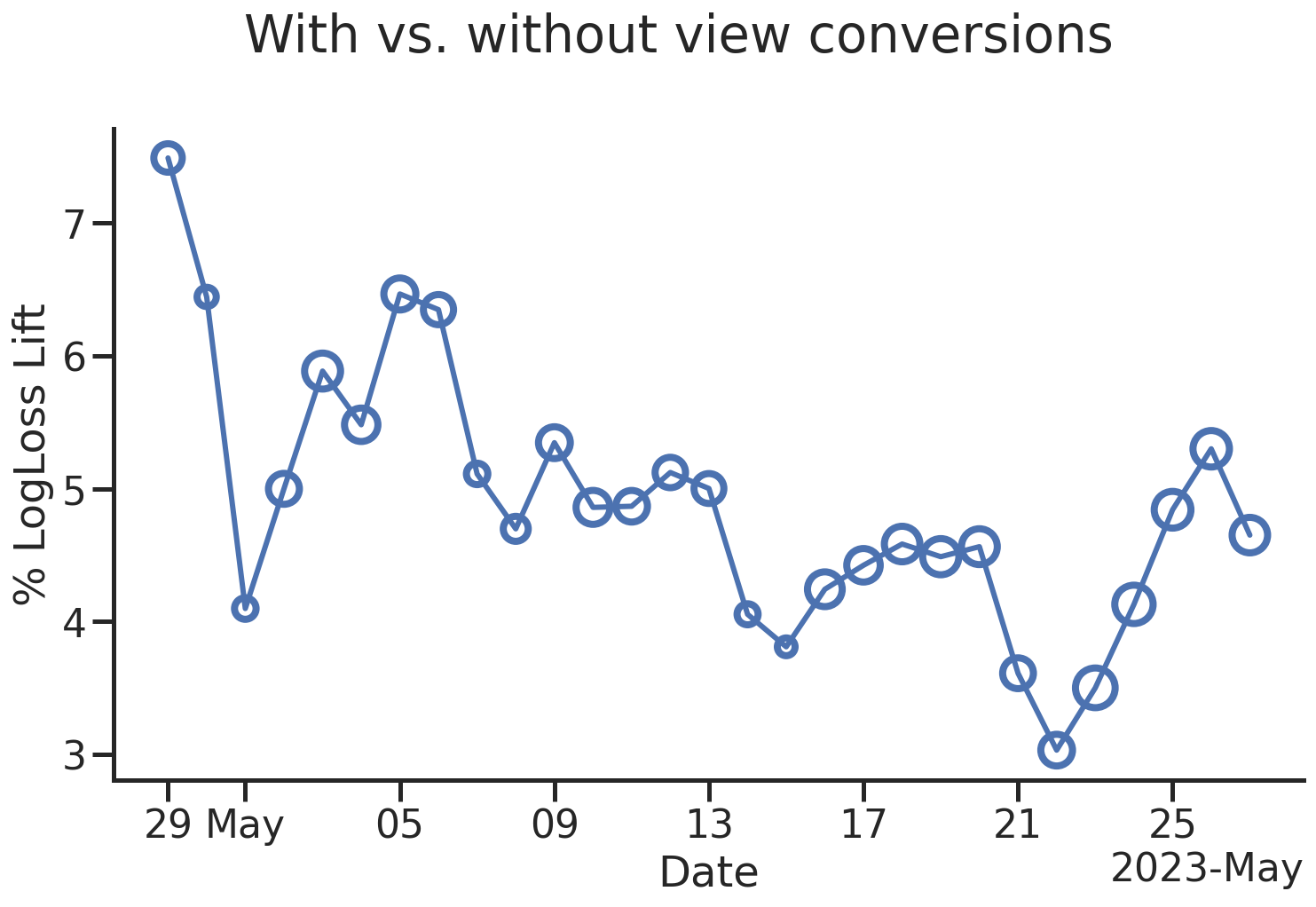}
    \caption{Daily \logloss difference, in percents, between a model that includes view-attributed conversions in its training set, and a model that does not. A positive difference means worse accuracy. Circle size is proportional to the size of the evaluation set.}
    \label{fig:view_conv_lift}
\end{figure}
}

\subsection{Auto-encoder evaluation}
Here we evaluated the desired properties of our auto-encoder using the metrics $\recloss_t$, $\recloss^R_t$, $\mathrm{Gen}_t$, and $\mathrm{Diff}_t$ from  Section \ref{sec:autoencoder_design}.

\begin{figure}[t]
    \centering
    \includegraphics[width=.9\columnwidth]{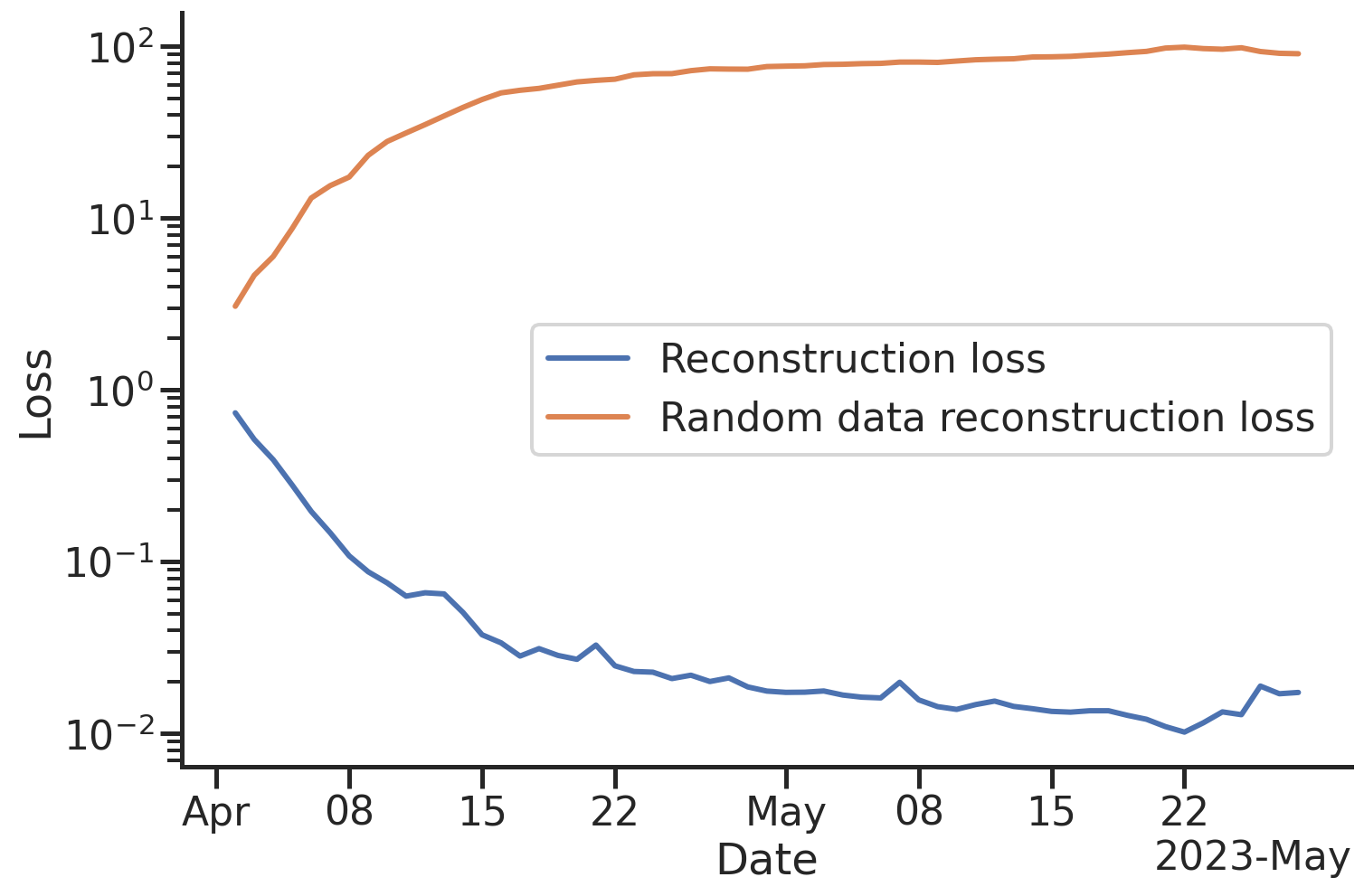}
    \caption{Reconstruction losses over real and random data, of an auto-encoder trained on a month worth of data.}
    \label{fig:real_vs_random}
\end{figure}

In order to validate that our auto-encoder learns meaningful patterns, rather than just the identity function, we measure $\recloss_t$ and $\recloss^R_t$ for an auto-encoder trained on a month worth of data. We plot the results in Figure \ref{fig:real_vs_random}. It is evident that as the model trains on more data, its performance on true data improves, on random data degrades, the reconstruction loss approaches zero, and the $\mathrm{RandRatio}_t$ approaches $10^{-4}$, as desired.

\begin{figure}
    \centering
    \includegraphics[width=.9\columnwidth]{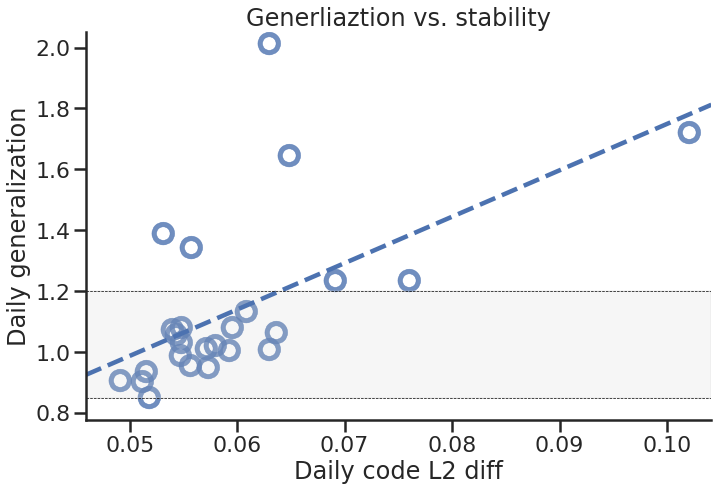}
    \caption{Daily generalization metric $\mathrm{Gen}_t$ (closer to 1 is better) versus daily stability metric $\mathrm{Diff}_t$ (lower is better).}
    \label{fig:generalization_stability}
\end{figure}

In order to verify that our auto-encoder is stable and generalizes well between consecutive days, we measure $\mathrm{Gen}_t$ and $\mathrm{Diff}_t$ over a few weeks worth of data. The results are plotted in Figure \ref{fig:generalization_stability}, where each point is the result of training on a day of data. As is apparent, most of the values of $\mathrm{Gen}_t$ are close to 1, which means that our auto-encoder often generalizes well. Moreover, as we claimed in Section \ref{sec:autoencoder_design}, generalization and stability are tightly coupled. Indeed, most of the points lie along a straight regression line, and there are no points where $\mathrm{Gen}_t$ is small but $\mathrm{Diff}_t$ is large. 

\subsection{CVR model offline lifts}
\begin{figure}
    \centering
    \includegraphics[width=.9\columnwidth]{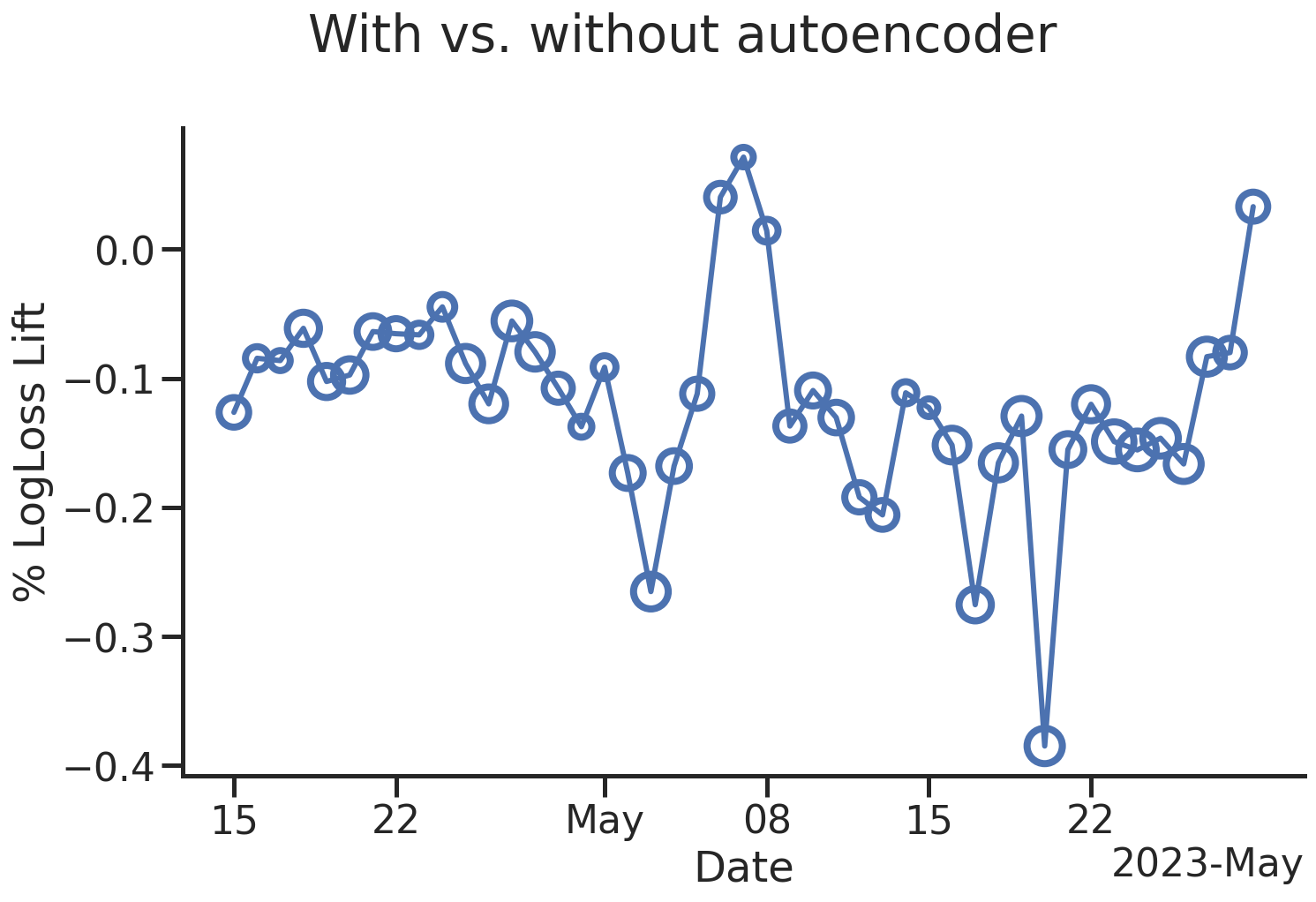}
    \caption{Daily \logloss difference (negative is better), in percents, between a model enriched by an auto-encoder, and the production model at the time. Marker size is proportional to the evaluation set size.}
    \label{fig:autoencoder_lift}
\end{figure}

We train a new CVR model using our auto-encoder based approach, and measure the \logloss difference between the new model and the production model at the time.  In contrast to a CVR model that is directly trained on view-attributed conversions, the auto-encoder based model has an improved \logloss, as is apparent from Figure \ref{fig:autoencoder_lift}.

\subsection{Incurred latency tests}
\begin{figure}
    \includegraphics[width=.9\columnwidth]{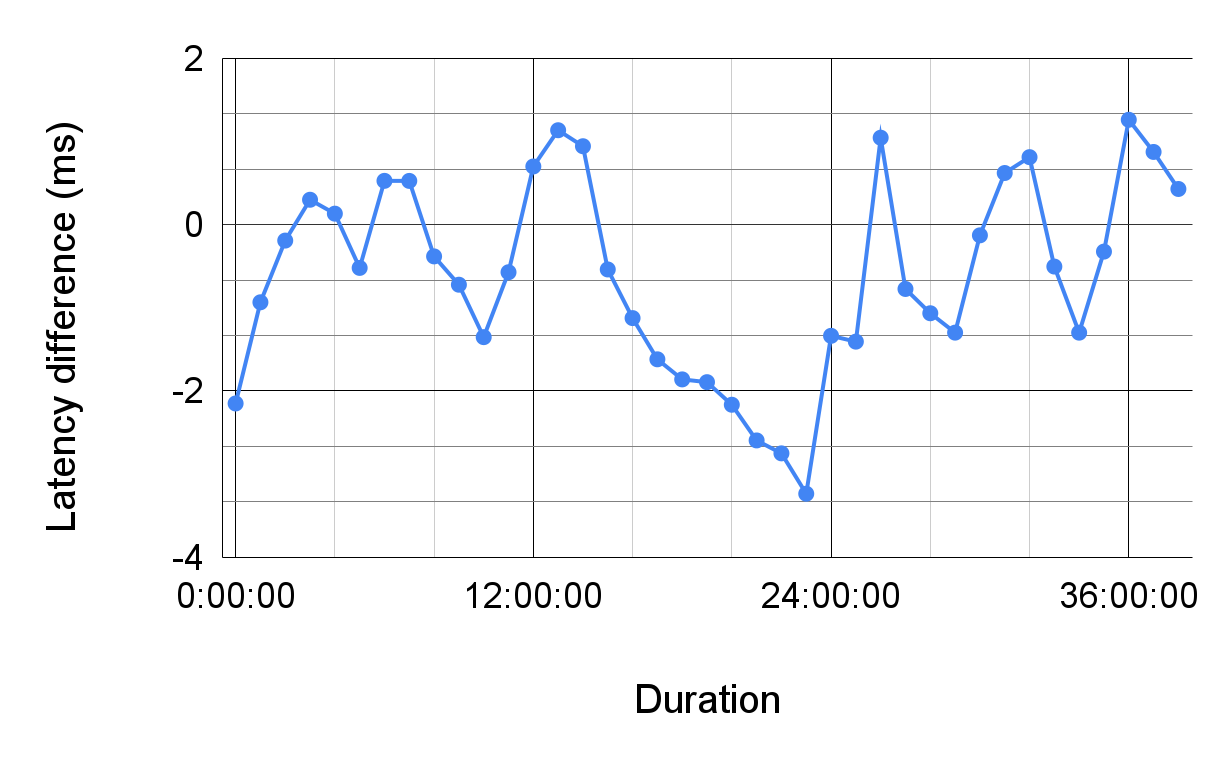}
    \caption{Average hourly latency difference, in milliseconds, between the new encoder-augmented model and the production model at the time. Negative values are better.}
    \label{fig:latency_diff}
\end{figure}

We deployed the new workflow to our production system by periodically loading the encoder that is trained with the model, and using it in ad auctions according to the formula in Equation \eqref{eq:modified_offset_model}. Since we use only high level ad taxonomy features, our ad auction computes the inner product $\langle \bv{W}, C(\Omega) \rangle$ only once for each ad category, and uses the cached results for all the ads participating in the auction. We compare the latency incurred by auctions where our new model is used to auctions with the production model at that time in Figure \ref{fig:latency_diff}. Indeed, the size of the model and the usage of high level features paid off - there is essentially no visible latency difference between the new and the product model at the time.

\subsection{Online A/B tests}

\paragraph{Setup}
A CVR model performance is measured along two axes - the revenue measured by the \textit{average cost per thousand impressions} (CPM), and the percentage of advertiser spend belonging to campaigns that attain advertiser cost per acquisition (CPA) goals. The objective is obtaining improvement in at least one metric without impairing the other. This is since a better model can either correct under-prediction, which results in lower bids and lower revenue, or over-prediction, which results in missing the advertiser CPA goal. To evaluate the online performance, we launched an online bucket serving 50\% of Yahoo native traffic and measured lifts of both metrics against the production model. The evaluation was conducted over a week and included billions of impressions. 

\paragraph{Results}
On average, the online test bucket showed $0.6$\% CPM lift and $0.67$\% revenue lift overall. Such a CPM (or revenue) lift translates to many millions of USD in yearly revenue once the solution is deployed to all traffic. The results also indicate a $+10.1$\% increase to the spend that is under oCPC influence, with a slightly larger percentage of spend belonging to campaigns hitting the CPA goal ($83.45$\% vs  $82.96$\% in test and control bucket receptively). In short, the results show that the new model increases the overall revenue while maintaining the advertiser's CPA goals. Following these positive results, the new model was fully deployed and serves 100\% of the Yahoo Gemini native traffic.

\section{Discussion and future work}
We described a framework for reducing the effect of data sparsity in CVR prediction by incorporating conversions that are considered un-labaled via self-supervised pre-training. Moreover, we adapted the basic idea to systems based on incrementally-trained factorization machines that are used in ad auctions with strict latency constraints. We have shown that even a simple auto-encoder with a small number of parameters already provides a significant improvement. Therefore, natural future directions, beyond adding more features, should aim to squeeze more information out of the un-labaled conversions by further improvements of model architecture and training techniques.

One direction is looking for better ways to incorporate the encoder's output into the factorization model without incurring a significant latency increase. Currently, we are using the random Fourier features technique as a `projector network'. It can be viewed as a shallow one-layer network whose activation is the cosine function, and its weights are random, rather than learned. However, the projector does not have to be a random one-layer network. It can be an arbitrary deep neural network with arbitrary activation functions and learnable weights. As long as this network is small enough to facilitate fast inference, it can potentially squeeze out more performance out of the encoder network. Moreover, the projector's incorporation into the factorization model does not have to be in the form of an additive term, but can be in the form of another latent vector. For example, instead modifying the score in Equation \eqref{eq:offset_score} as in \eqref{eq:modified_offset_model}, it can be alternatively modified as
\[
f(\Omega) = \langle \bv{U}, \bv{A}, P(\code(\Omega)) \rangle + \sum_{i\in S}w_i x_i + b,
\]
where $\langle \bv{x}, \bv{y}, \bv{z} \rangle = \sum_{i=1}^n x_i y_i z_i$ denotes a "triple" inner product, and $P(\cdot)$ is the projector network whose output dimension is equal to the dimension of the user and ad vectors. 

Another prominent direction is using more advanced self-supervised learning techniques, such as masked auto-encoders \cite{masked_autoenc}, or even drifting away from the auto-encoder framework and use techniques such as Barlow Twins \cite{barlow_twins}. Another improvement can potentially come from better model architectures, such as using small transformers \cite{attention_all_you_need} to process feature embedding vectors.

\bibliographystyle{plain}
\bibliography{autoencoder}

\end{document}